\documentstyle[12pt]{article}
\advance \topmargin by -\headheight
\advance \topmargin by -\headsep

\evensidemargin \oddsidemargin
\begin{document}

\topmargin -.6in

%
%
\def\rf#1{(\ref{eq:#1})}
\def\lab#1{\label{eq:#1}}
\def\nonu{\nonumber}
\def\br{\begin{eqnarray}}
\def\er{\end{eqnarray}}
\def\be{\begin{equation}}
\def\ee{\end{equation}}
\def\foot#1{\footnotemark\footnotetext{#1}}
\def\lb{\lbrack}
\def\rb{\rbrack}
\def\llangle{\left\langle}
\def\rrangle{\right\rangle}
\def\blangle{\Bigl\langle}
\def\brangle{\Bigr\rangle}
\def\llbrack{\left\lbrack}
\def\rrbrack{\right\rbrack}
\def\lcurl{\left\{}
\def\rcurl{\right\}}
\def\({\left(}
\def\){\right)}
\def\v{\vert}
\def\bv{\bigm\vert}
\def\Bgv{\;\Bigg\vert}
\def\bgv{\bigg\vert}
\def\lskip{\vskip\baselineskip\vskip-\parskip\noindent}
\relax

\def\tr{\mathop{\rm tr}}
\def\Tr{\mathop{\rm Tr}}
\def\partder#1#2{{{\partial #1}\over{\partial #2}}}
\def\funcder#1#2{{{\delta #1}\over{\delta #2}}}
\def\me#1#2{\left\langle #1\right|\left. #2 \right\rangle}
\def\a{\alpha}
\def\b{\beta}
\def\d{\delta}
\def\D{\Delta}
\def\eps{\epsilon}
\def\vareps{\varepsilon}
\def\g{\gamma}
\def\G{\Gamma}
\def\grad{\nabla}
\def\h{{1\over 2}}
\def\l{\lambda}
\def\L{\Lambda}
\def\m{\mu}
\def\n{\nu}
\def\o{\over}
\def\om{\omega}
\def\O{\Omega}
\def\p{\phi}
\def\P{\Phi}
\def\vp{\varphi}
\def\va{\vartheta}
\def\pa{\partial}
\def\pr{\prime}
\def\qb{{\bar q}}
\def\qt{{\tilde q}}
\def\ra{\rightarrow}
\def\s{\sigma}
\def\S{\Sigma}
\def\t{\tau}
\def\th{\theta}
\def\Th{\Theta}
\def\ti{\tilde}
\def\wti{\widetilde}
\def\veps{\varepsilon}
\def\tg{\bigtriangleup}
%
\def\lie{{\cal G}}
\def\dlie{{\cal G}^{\ast}}
\def\elie{{\widetilde \lie}}
\def\edlie{{\elie}^{\ast}}
\def\hatlie{{\hat {\lie}}}
\def\hlie{{\cal H}}
\def\wlie{{\widetilde \lie}}
\def\sur{$SU(r+1)$}
\def\spr{$Sp(r)$}
\def\soe{$SO(2r)$}
\def\sod{$SO(2r+1)$}
\def\gtwo{$G_2$}
\def\ffour{$F_4$}
\def\esix{$E_6$}
\def\eeight{$E_8$}
\def\sw{w_{\infty}}
\def\bw{W_{\infty}}
\def\osw{w_{1+ \infty}}
\def\obw{W_{1+\infty}}

\def\rlx{\relax\leavevmode}
\def\inbar{\vrule height1.5ex width.4pt depth0pt}
\def\IZ{\rlx\hbox{\sf Z\kern-.4em Z}}
\def\IR{\rlx\hbox{\rm I\kern-.18em R}}
%
%
\def\mark{\noindent{\bf Remark.}\quad}
\def\prop{\noindent{\bf Proposition.}\quad}
\def\proof{\noindent{\bf Proof.}\quad}
\newcommand{\nit}{\noindent}
\newcommand{\ct}[1]{\cite{#1}}
\newcommand{\bi}[1]{\bibitem{#1}}
\newtheorem{theor}{Theorem}[section]
%
%
\def\PRL#1#2#3{{\sl Phys. Rev. Lett.} {\bf#1} (#2) #3}
\def\NPB#1#2#3{{\sl Nucl. Phys.} {\bf B#1} (#2) #3}
\def\NPBFS#1#2#3#4{{\sl Nucl. Phys.} {\bf B#2} [FS#1] (#3) #4}
\def\CMP#1#2#3{{\sl Commun. Math. Phys.} {\bf #1} (#2) #3}
\def\PRD#1#2#3{{\sl Phys. Rev.} {\bf D#1} (#2) #3}
\def\PLA#1#2#3{{\sl Phys. Lett.} {\bf #1A} (#2) #3}
\def\PLB#1#2#3{{\sl Phys. Lett.} {\bf #1B} (#2) #3}
\def\JMP#1#2#3{{\sl J. Math. Phys.} {\bf #1} (#2) #3}
\def\PTP#1#2#3{{\sl Prog. Theor. Phys.} {\bf #1} (#2) #3}
\def\SPTP#1#2#3{{\sl Suppl. Prog. Theor. Phys.} {\bf #1} (#2) #3}
\def\AoP#1#2#3{{\sl Ann. of Phys.} {\bf #1} (#2) #3}
\def\PNAS#1#2#3{{\sl Proc. Natl. Acad. Sci. USA} {\bf #1} (#2) #3}
\def\RMP#1#2#3{{\sl Rev. Mod. Phys.} {\bf #1} (#2) #3}
\def\PR#1#2#3{{\sl Phys. Reports} {\bf #1} (#2) #3}
\def\AoM#1#2#3{{\sl Ann. of Math.} {\bf #1} (#2) #3}
\def\UMN#1#2#3{{\sl Usp. Mat. Nauk} {\bf #1} (#2) #3}
\def\FAP#1#2#3{{\sl Funkt. Anal. Prilozheniya} {\bf #1} (#2) #3}
\def\FAaIA#1#2#3{{\sl Functional Analysis and Its Application} {\bf #1} (#2)
#3}
\def\BAMS#1#2#3{{\sl Bull. Am. Math. Soc.} {\bf #1} (#2) #3}
\def\TAMS#1#2#3{{\sl Trans. Am. Math. Soc.} {\bf #1} (#2) #3}
\def\Invm#1#2#3{{\sl Invent. Math.} {\bf #1} (#2) #3}
\def\LMP#1#2#3{{\sl Letters in Math. Phys.} {\bf #1} (#2) #3}
\def\IJMPA#1#2#3{{\sl Int. J. Mod. Phys.} {\bf A#1} (#2) #3}
\def\AdM#1#2#3{{\sl Advances in Math.} {\bf #1} (#2) #3}
\def\RMaP#1#2#3{{\sl Reports on Math. Phys.} {\bf #1} (#2) #3}
\def\IJM#1#2#3{{\sl Ill. J. Math.} {\bf #1} (#2) #3}
\def\APP#1#2#3{{\sl Acta Phys. Polon.} {\bf #1} (#2) #3}
\def\TMP#1#2#3{{\sl Theor. Mat. Phys.} {\bf #1} (#2) #3}
\def\JPA#1#2#3{{\sl J. Physics} {\bf A#1} (#2) #3}
\def\JSM#1#2#3{{\sl J. Soviet Math.} {\bf #1} (#2) #3}
\def\MPLA#1#2#3{{\sl Mod. Phys. Lett.} {\bf A#1} (#2) #3}
\def\JETP#1#2#3{{\sl Sov. Phys. JETP} {\bf #1} (#2) #3}
\def\JETPL#1#2#3{{\sl  Sov. Phys. JETP Lett.} {\bf #1} (#2) #3}
%

\begin{titlepage}
\vspace*{-1cm}
\noindent
April, 1993 \hfill{IFT-P/020/93}\\
\phantom{bla}
\hfill{UICHEP-TH/93-3} \\
\phantom{bla}
\hfill{hep-th/9304080}
\\
\vskip .3in
\begin{center}
{\large\bf Construction of  Affine and Conformal Affine Toda Solitons by
Hirota's method \footnotemark
\footnotetext{Talk presented at the VII J.A. Swieca Summer School, Section:
 Particles and Fields,  Campos do Jord\~ao - Brasil - January/93}}
\end{center}

\normalsize
\vskip .4in

\begin{center}
{ H. Aratyn\footnotemark
\footnotetext{Work supported in part by U.S. Department of Energy,
contract DE-FG02-84ER40173 and by NSF, grant no. INT-9015799}}

\par \vskip .1in \noindent
Department of Physics \\
University of Illinois at Chicago\\
801 W. Taylor St.\\
Chicago, Illinois 60607-7059\\
\par \vskip .3in

\end{center}

\begin{center}
{C.P. Constantinidis\footnotemark\footnotetext{Supported by Fapesp},
L.A. Ferreira\footnotemark
\footnotetext{Work supported in part by CNPq}}, J.F. Gomes$^{\,4}$ and
A.H. Zimerman$^{\,4}$

\par \vskip .1in \noindent
Instituto de F\'{\i}sica Te\'{o}rica-UNESP\\
Rua Pamplona 145\\
01405-900 S\~{a}o Paulo, Brazil
\par \vskip .3in

\end{center}

\begin{center}
{\large {\bf ABSTRACT}}\\
\end{center}
\par \vskip .3in \noindent

In this talk we report some results about the construction of soliton solutions
for the Affine and Conformal Affine Toda models using the Hirota's method.  We
obtain new classes of solitons connected to the degeneracies of the Cartan
 matrix eigenvalues as well as to some particular features of the recursive
 scheme developed here. We obtain an universal mass  formula for all those
 solitons. The examples of $SU(6)$ and $Sp(3)$ are discussed in some detail.

\end{titlepage}

\section{Introduction}
\setcounter{equation}{0}

One of the main motivations for searching for solitons solutions in the Toda
models comes from the interest in generalizing the well known soliton solutions
in the celebrated sine-Gordon model which equation of motion in two dimensions
reads:
\be
\pa_{-} \pa_{+} \vp = \, {q \o \qb} \sin (2 \qb \vp)
\lab{sgordon}
\ee

In this case the vacuum configuration is degenerate and this is
responsible for the topological nature of the soliton solutions.

{}From the point of view of Toda theories the sine-Gordon model
can be considered as the simplest example of the Affine Toda (AT)
model whose equations of motion are given by:
\be
\pa_{-} \pa_{+} \vp^a = \, {1 \o \qb}\( q^{a} e^{\qb K_{ab} \vp^b}
- \,q^{0} l^{\psi}_{a} e^{- \qb K_{\psi b} \vp^b } \)
\lab{todaaf}
\ee
where $K_{ab}=2 \a_a.\a_b/{\a_b^2}$ is the Cartan Matrix of $\lie$,
$a,b=1,...$, ${\rm rank} \,\lie\equiv r$, $\psi$ is the highest root of $\lie$,
$K_{\psi b}=2 \psi .\a_b/{\a_b^2}$, $l^{\psi}_{a}$ are positive integers
appearing in the expansion ${\psi \over
\psi^{2}} = l^{\psi}_{a} {\a_{a} \over \a^{2}_{a}}$, where $\a_a$ are the
simple roots of $\lie$ and $q^a$, $q^0$ and $\qb$ are coupling constants.
Indeed, the sine-Gordon model equation of motion is obtained if
one considers $\lie$ as $su(2)$. In this case $K_{ab}=2$ and
$\vp$ has only one component. The algebraic structure of the
Affine Toda models is given by a loop algebra associated to $\lie$.

Another hierarchy of Toda theories is obtained from the AT ones by adding
two extra fields $\eta$ and $\nu$ so that \ct{AFGZ,BB}:
\br
\pa_{-} \pa_{+} \vp^a &=& \, {1 \o \qb}\( q^{a} e^{\qb K_{ab} \vp^b}
- \,q^{0} l^{\psi}_{a}  e^{- \qb K_{\psi b} \vp^b } \) e^{\qb \eta}
\lab{todaone} \\
\pa_{-} \pa_{+} \eta &=& 0 \lab{todatwo} \\
\pa_{-} \pa_{+} \nu &=&   {2 \o \psi^2}
\, {q^{0}\o \qb} e^{-\qb K_{\psi b} \vp^{b}} e^{\qb \eta}
\lab{todathree}
\er
are the equations of motion of the so-called Conformal Affine
Toda model (CAT), related to Kac-Moody algebras.
These equations are invariant under conformal transformations:
\br
x_{+} \rightarrow \tilde{x}_{+} = f(x_{+})
 \, \, \, , \, \, \, \,
x_{-} \rightarrow \tilde{x}_{-} = g(x_{-}) \lab{ge}
\er
and
\br
e^{-\vp^a (x_+,x_-)} &\to&
e^{-\tilde{\vp}^a(\tilde{x}_+,\tilde{x}_-)} = e^{-\vp^a (x_+,x_-)}
\lab{fi} \\
e^{-\nu (x_+,x_-)} &\to&
e^{-\tilde{\nu}(\tilde{x}_+,\tilde{x}_-)}= ({df \over dx_+})^{B}
({dg \over dx_-})^{B} e^{-\nu (x_+,x_-)} \lab{ni}\\
e^{-\eta (x_+,x_-)} &\to&
e^{-\tilde{\eta}(\tilde{x}_+,\tilde{x}_-)} =({df \over dx_+})^{{1\over \qb}}
({dg \over dx_-})^{{1 \o \qb}} e^{-\eta (x_+,x_-)}   \lab{mi}
\er
where $f$ and $g$ are analytic functions and $B$ is an arbitrary constant.
Therefore $e^{\varphi^a}$ are scalars under conformal transformations
and $e^{ \nu} $ can also be arranged to be a scalar by setting $B=0$ \ct{CFGZ}.
The AT models, on the other hand, are not invariant under conformal
 transformations.

{}From \rf{todaone} it is noticed  that the AT model can be
understood as a CAT model when the conformal symmetry is in some
sense gauge fixed \ct{CFGZ}. Making a suitable choice of
transformations :
\be
f^{\prime}(x_{+}) = e^{\eta_{+}(x_{+})} \, \, \, , \, \, \,
g^{\prime}(x_{-}) = e^{\eta_{-}(x_{-})}
\label{etatransf}
\ee
where $\eta_{\pm}(x_{\pm})$ are solutions of the $\eta$ field, i.e., $\eta
(x_{+},x_{-}) = \eta_{+}(x_{+}) + \eta_{-}(x_{-})$, the CAT
model equations can be written as:
\be
\tilde \pa_{-}  \tilde \pa_{+} \tilde \varphi^a = \,
q^{a}e^{K_{ab} \tilde \varphi^{b}} - \,l^{\psi}_{a}q^{0}e^{-
K_{\psi b} \tilde \varphi^{b}}
\lab{decoup1}
\ee
\be
\tilde \pa_{-} \tilde \pa_{+} \tilde \nu =   {2 \o \psi^2}
\, q^{0}e^{- K_{\psi b} \tilde \varphi^{b}}  \label{eq:decoup2}
\ee
and the new space time coordinates are determined in terms of the old ones
through the given solution of $\eta$, i.e. $\tilde x_{+} = \int^{x_{+}}
d{y}_{+} e^{\eta_{+} ({y}_{+})}$, $\tilde x_{-} = \int^{x_{-}} d{y}_{-}
e^{\eta_{-} ({ y}_{-})}$. Observe that in these new coordinates
equations \rf{decoup1} are the AT equations of motion.

The equations of motion of both hierarchies can be written as
a zero curvature condition:
\be
\pa_{+} A_{-} - \pa_{-} A_{+} + \lb A_{+} , A_{-} \rb = 0
\ee
In the case of AT models the gauge potentials $A_{\pm}$ lie on a loop algebra
 associated to $\lie$, whilst for the CAT models they lie on a Affine Kac-Moody
algebra ${\hat \lie}$.

The solitonic character of the AT model can be observed from the potential
of the theory. Introducing, for convenience, the following vector:
\be
\vp \equiv \sum_{a=1}^r {2 \a_a \o \a_a^2} \vp^a
\lab{vectorphi}
\ee
the potential can then be written as
\be
U(\vp , \eta ) = \sum_{j=0}^r {4 q^j\o \a_j^2} e^{\qb \left( \a_j.\vp +
\eta \right)}
\lab{potential2}
\ee
where $\a_0 = - \psi$ is denoted as the extra simple root of the affine
Kac-Moody algebra ${\hat \lie}$.

This potential  is invariant under the transformation
\be
\vp \rightarrow \vp + {2 \pi i \o \qb} \mu^{v} \, \, \, ; \, \, \,
\eta \rightarrow \eta + {2 \pi i \o \qb} n
\lab{degvacua}
\ee
where $\mu^{v}$ is a coweight of $\lie$, i.e. $\mu^v = \sum_{a=1}^r
m_a {2 \lambda_a \o \a_a^2}$ where $m_a$ and $n$ are integers, and $\lambda_a$
are the fundamental weights of $\lie$.
For $\qb$ purely imaginary the vacuum is infinitely degenerate,
corresponding to the coweight lattice of $\lie$. This generalizes
to any simple Lie algebra the degenerate vacua of the sine Gordon
model where the minima of the potential are identified with co-weight
lattice of $SU(2)$.  The degeneracy shown above indicates the existence
of topological solitons in the AT and CAT models.

\section{Mass and Charge of the solitons}
\setcounter{equation}{0}

The energy-momentum tensor of the CAT model can be made traceless due to
scale invariance:
\be
\Theta_{\rho \sigma}^{\rm CAT} = \Theta_{\rho \sigma}^{\rm CAN}
 - \qb \(\pa_{\rho}
\pa_{\sigma} - g_{\rho \sigma} \pa^2 \) \( \sum_{a=1}^r {2\o \a^2_a} \vp^a + h
\nu\) \lab{cattensor}
\ee
where $\Theta^{\rm CAN}$ is the canonical energy-momentum tensor, and
 $h$ is the Coxeter number of $\lie$. It is easy to  verify
that $\Theta_{\rho}^{\rho {\rm CAT}} = 0$ and
$\Theta_{\rho}^{\rho {\rm CAN}} = 2 U(\vp ,\eta)$.

Consider a soliton type solution which can be put at rest at
some Lorentz frame. Then the energy of the solution should be
the mass of the soliton. Such mass should be proportional to
some mass scale of the theory. Due to conformal (scale) invariance
such mass scale does not exist in the CAT model and therefore the
solitons are massless. For the AT model, however, such argument does
not apply and the solitons are massive. In this case:
\be
\Theta_{\rho \sigma}^{\rm AT} =
\Theta_{\rho \sigma}^{\rm CAN}\mid_{\eta = 0}
\lab{emcat/at}
\ee

One then observes from \rf{cattensor} and \rf{emcat/at} that the contribution
to the soliton mass in the AT models comes from a total divergence term, which
involves the CAT field $\nu$.
Indeed, denoting by $M$ the soliton mass and $v$ its velocity (in
units of light velocity) one gets:
\br
{M v \o {\sqrt{1 - v^2}}} &=& \int^{\infty}_{-\infty} dx \, \Theta^{AT}_{01}
\nonumber\\
&=& \qb \int^{\infty}_{-\infty} dx \pa_{x}\pa_{t} \( \sum_{a=1}^r {2\o
\a^2_a} \vp^a + h \nu\) \lab{divergence}\\
&=& \qb \pa_{t} \( \sum_{a=1}^r {2\o \a^2_a} \vp^a +
h\nu\)\mid_{-\infty}^{\infty} \nonumber
\er
Observe that this is a universal formula which can be used to determine  the
masses of the solitons \cite{Olive,ACFGZ}.

The topological charges of  the solitons can still be
 introduced as :
\be
Q \equiv {\qt \o 2 \pi} \int_{-\infty}^{\infty} dx \pa_x \vp
=  {\qt \o 2 \pi}\left( \vp (\infty ) - \vp (-\infty )\right)
\lab{charge}
\ee
where $i\qt \equiv \qb$ and it will be a vector in the root space.

\section{Construction of solitons: Hirota's Method}
\setcounter{equation}{0}

In order to follow Hirota's procedure \ct{Hirota} of constructing soliton
solutions for nonlinear systems  we introduce the $\tau$ functions as
\ct{hollo,CFGZ,MM,ACFGZ}:
\be
\varphi^a ={1\o \qb}\( - \ln {\tau_a \over \tau_0^{l_a^{\psi}}} + \va_a \)
\;\;\;\;\;\;\;\;
\nu ={1\o\qb} {2 \over {\psi}^2} \left( \sigma - \ln \tau_0 \right)
\lab{phinu}
\ee
with $a=1,2,...,$ rank-$\lie$ and $\va_a$ are constants depending on the
coupling constants $q^0$, $q^a$ and $\qb$ (see \ct{ACFGZ} for more details).

For $\eta =0$ the CAT equations \rf{todaone} and \rf{todathree} can
be decoupled into
\br
\tg (\tau_j) &=& \beta l_j^{\psi} \left(  1 -\prod_{k=0}^{rank \lie}
\tau_k^{-K_{jk}} \right) \lab{tau1}\\
\pa_{+} \pa_{-} \sigma &=& \beta
\lab{tau2}
\er
with
\be
\bigtriangleup (F) \equiv \pa_{+} \pa_{-} \ln F = {\pa_{+} \pa_{-} F \over F}
- {\pa_{+} F \pa_{-} F \over F^2}
\lab{triangle}
\ee
and
\be
\beta = {q^j \over l_j^{\psi}} e^{K_{jk}\va_k} \, \, \, \, \, \, \, \,
\mbox{for any $j=0,1,\ldots,r$}
\lab{defbeta}
\ee
is a constant independent of the index $j$ \ct{CFGZ}. Now
$K_{jk}$ is the extended Cartan matrix of ${\hat \lie}$ and in the
calculation it was used the fact that $l^{\psi}_{j}$,
$j=0,1,2,...,$ rank-$\lie$, with $l_0^{\psi}=1$, constitute a
null vector of the extended Cartan matrix, i.e.,
$\sum_{j=0}^{r}K_{ij}l_j^{\psi}= 0$.

The solution to \rf{tau2} is given by
\be
\sigma (x_{+}, x_{-}) = \beta x_{+} x_{-} + F(x_{+}) + G(x_{-})
\lab{defsigma}
\ee
with $F$ and $G$ being arbitrary functions. The solution for the
$\tau$'s, in Hirota's method spirit, is constructed using the
following expansion:
\be
\tau_i = 1 + \epsilon \tau^{(1)}_i + ... + \epsilon^{N_i} \tau^{(N_i)}_i
\lab{Nsoliton}
\ee
with ansatz \ct{hollo}:
\be
\t^{(n)}_i = \d^{(n)}_i \; e^{n \Gamma}
\lab{ansatz}
\ee

\be
\Gamma =  \gamma ( x - v t ) + \xi
\lab{Gamma}
\ee
where $\d^{(n)}_i$ are constant vectors to be found from Hirota's equations,
$\gamma$ and $\xi$ are parameters of the solution. One can show
that highest orders of $\tau$'s constitute a null vector of the
extended Cartan matrix \ct{ACFGZ}
\be
K_{ij}\, N_j = 0
\lab{hpower}
\ee
Therefore $N_i = \kappa\, l_i^{\psi}$, where $\kappa$ is some positive integer.

Substituting expansion \rf{Nsoliton} in \rf{todaone} one
finds in first order in $\epsilon$:
\be
L_{ij} \delta^{(1)}_j = \lambda \delta^{(1)}_i
\lab{nld}
\ee
where $L_{ij} = l_i^{\psi} K_{ij}$, and $\l = {\gamma^2 (1 - v^2)\o 4 \beta}$.
Therefore the parameters of the solution are restricted by the possible
eigenvalues of the matrix $L_{ij}$.

The higher $\delta$'s are determined from $\delta^{(1)}$ recursively through
\ct{ACFGZ} :
\be
\d^{(n)}_i = {S^{(n)}}^{-1}_{ij} V^{(n-1)}_j \lab{dfromv}
\ee
with
\be
S^{(n)} \equiv L - n^2 \l
\lab{dfromv2}
\ee
and
$V_j^{(n-1)} \equiv$ function of $\delta$ 's of order $(n-1)$ or
less.

Higher $\delta$'s are uniquely determined by $\delta^{(1)}$
except for the cases where $L$ has two eigenvalues $\l$ and
$\l$' satisfying
\be
\l = n^2 \l^{\prime} \, \, \, \, \mbox{\rm for some integer $n$}
\lab{2ndtyp}
\ee
This kind of degeneracy appears only for $SU(6p)$ and $Sp(3p)$,
with $p$ a positive integer.

Once the $\delta$'s  are determined one is able to write $\vp$ explicitly
and so the soliton masses are calculated using the universal formula
\rf{divergence}:
\be
M =  {4 h \kappa  \o \psi^2} m  \sqrt{ \l}
\lab{massfinal}
\ee
where $m \equiv \sqrt{\b}$, $h$ is the Coxeter number and
$\kappa$ is the integer given by $N_i = \kappa\, l_i^{\psi}$.

The procedure described above was used to obtain the solitons solutions of the
CAT and AT models associated to any simple Lie algebra \ct{ACFGZ}. For
 simplicity we have chosen to discuss here the examples of $SU(6)$ and $Sp(3)$
 which possess all the features of the procedure.

\section{The Example of $SU(6)$ }
\setcounter{equation}{0}

The $L$ matrix in
this case coincides with the extended Cartan matrix since
$l^{\psi}_i = 1$ for any $i=0,1, \ldots , 5$ and is written as
\be
K= \left( \begin{array}{rrrrrr}
2  & -1 & 0 & 0 & 0 & -1\\
-1 & 2 & -1 & 0 & 0 & 0 \\
0  & -1 & 2 & -1 & 0 & 0 \\
0 & 0 & -1 & 2 & -1 & 0  \\
0 & 0 & 0 & -1 & 2 & -1 \\
-1 & 0 & 0 & 0 & -1 & 2 \end{array} \right)
\label{cartsu6}
\ee
and Hirota's equation \rf{tau1} reads
\be
\tau_j^{2}\tg (\tau_{j}) =  \b \( \tau_j^{2} -
\tau_{j+1} \tau_{j-1}\right) \, \, \, \, \, \, \, \mbox{for
$j=0,1,2,\ldots,r$}
\lab{sunhiro}
\ee
where, due to the periodicity of the extended Dynkin diagram, it is understood
that $\tau_{j+6} = \tau_{j}$.

Remember that the $\delta^{(1)}$'s are obtained from the first
order term of Hirota's expansion as the eigenvectors of $L$. The eigenvalues of
 $L$ are:
\be
\l_j \,= \, 4 \sin^2 \( {j \pi \o 6 }\)  \, \, \, \, \, \, \, \mbox{for
$j=0, 1,2,\ldots,5$}
\lab{su6eigenv}
\ee
or $\l = (0,1,3,4,3,1)$ and the corresponding eigenvectors are
\br
v_k &=& 1 \, \, \, \, \mbox{for $\l = 0$}  \nonu \\
v_{\lb 1 \rb\, k} &=& \exp\({2\pi i  k \o 6}\) \quad;\quad
v_{\lb 2 \rb\, k} =\exp\(-{2\pi i  k \o 6}\) \, \, \,
\, \, \, \, \mbox{for $\l = 1$} \nonu \\
v_{\lb 1 \rb\, k} &=& \exp\({2\pi i  k \o 3}\) \quad;\quad
v_{\lb 2 \rb\, k} =\exp\(-{2\pi i  k \o 3}\) \, \, \,
\, \, \, \, \mbox{for $\l = 3$} \nonu \\
v_k &=& (-1)^k \quad \mbox{for $\l = 4$}  \nonu \\
\lab{su6vectors}
\er
where $k =0, 1,2,\ldots,5$

The function $V^{(n-1)}$, appearing in \rf{dfromv}, is obtained from
 \rf{sunhiro} and for the algebra which is being considered has the form :
\be
V^{(n-1)}_j = - \sum_{l=1}^n \( \( 1 -  \l \( n^2 - 3n l +2 l^2\) \) \d_j^{(l)}
\d_j^{(n-l)}   - \d_{j+1}^{(n-l)} \d_{j-1}^{(l)} \)
\lab{sunv}
\ee

For $\l = 0$ one gets a trivial solution where all $\vp$'s are
constant. For $\l = 4$, $\d_j^{(1)} = (-1)^j$ and $V_j^{(1)}=0$.
The series truncates at first order, so the $\tau$ function is
given by:
\be
\tau_j = 1 + (-1)^j e^{\Gamma}
\lab{taul4}
\ee
Substituting \rf{taul4} into \rf{phinu} one gets the explicit
expression for the $\vp$'s:
\be
\varphi^a ={1\o \qb}\( - \ln \({1 + (-1)^a e^{\Gamma} \over
1 + e^{\Gamma}}\) + \va_a \)
\ee
and the mass is, according to \rf{massfinal} $M={48 \o \psi^2}m$.

When $\l=3$ there is a degeneracy and $\d^{(1)}$ must be a linear
combination of its corresponding eigenvectors. The recurrence method
furnishes:
\be
\t_j = 1 + \(y_1 \exp\({2\pi i j \o 3}\) +
y_2 \exp\(-{2\pi i j \o 3}\) \) e^\G +  {1\o 4}y_1 y_2 e^{2\G}
\lab{su6tau3}
\ee
where $y_1$ and $y_2$ are free parameters and the masses are:
\be
M_l^{\l =3} ={24 \sqrt{3}\o{\psi^2}}m \, \, \, \, \,
\mbox{\rm for $y_1=0$ or $y_2=0$}
\ee
\be
M_2^{\l =3} ={48 \sqrt{3}\o{\psi^2}}m \, \, \, \, \,
\mbox{\rm for $y_1,y_2 \neq 0$}
\ee

The last eigenvalue, $\l=1$, is also degenerate. Following the
same steps as for the case $\l=3$, $\d^{(1)}$ must be a linear combination
of the corresponding eigenvectors. However, in this case it will
appear the second type degeneracy \rf{2ndtyp}, since $4$ is also
an eigenvalue: $4=2^2.1$. Due to this fact $\d^{(2)}$ is not
uniquely determined from $\d^{(1)}$. The $\tau$ function is now
written as:
\br
\t_j &=& 1 + \(y_1 \exp\({i \pi j \o 3}\) +
y_2 \exp\(-{i \pi j \o 3}\) \) e^\G + \( {3 \o 4}  y_1 y_2 + (-1)^j z \)
e^{2\G}\nonu \\
 &+& {z\o 3}  (-1)^j \( y_1 \exp\({i \pi j\o 3}\) +y_2
\exp\(-{i \pi j\o 3}\)\) e^{3 \G}
+(-1)^j {z y_1 y_2\o 12} e^{4 \G}
\lab{suntaugalois}
\er
and the masses are:
\br
M_1^{\l =1} &=&{24 \o{\psi^2}}m \, \, \, \, \,
\mbox{\rm for $z=0$ and $y_1=0$ or $y_2=0$}\\
M_2^{\l =1} &=&{48 \o{\psi^2}}m  \, \, \, \, \,
\mbox{\rm for $z=0$ and $y_1,y_2 \neq 0$
or $z \neq 0$ and $y_1=y_2=0$}\\
M_3^{\l =1} &=&{72 \o{\psi^2}}m \, \, \, \, \,
\mbox{\rm for $z \neq 0$ and $y_1=0$ or $y_2=0$}\\
M_4^{\l =1} &=&{96 \o{\psi^2}}m \, \, \, \, \,
\mbox{\rm for $z, y_1, y_2 \neq 0$}
\er

\section{The Example of $Sp(3)$}
\setcounter{equation}{0}

The Cartan matrix is  given by
\be
K= \left( \begin{array}{rrrr}
2 & -2 & 0 & 0  \\
-1 & 2 & -1 & 0  \\
0 & -1 & 2 & -1 \\
0 & 0 & -2 & 2 \end{array} \right)
\ee
$l_i^{\psi} =1$ for $i=0,1,2,3$, and yields the following  Hirota's
equations
\br
\tau_0^{2} \tg (\tau_0) &=& \beta \left( \tau_0^{2} -
\tau_1^{2}\right) \nonu\\
\tau_1^{2}\tg (\tau_{1}) &=& \beta \left( \tau_1^{2} -
\tau_{2} \tau_{0}\right) \nonu\\
\tau_2^{2}\tg (\tau_{2}) &=& \beta \left( \tau_2^{2} -
\tau_{1} \tau_{3}\right) \nonu\\
\tau_3^{2}\tg (\tau_3) &=& \beta \left(\tau_3^{2} -  \tau_{2}^{2}\right)
\lab{hirospr}
\er

One notices that such equations can be obtained from Hirota's eqs. \rf{sunhiro}
for $SU(6)$ by making the identifications $\tau_5 \equiv \tau_1$ and
$\tau_4 \equiv \tau_2$. Therefore, all solutions of \rf{sunhiro} which are
invariant  under the interchanges  $\tau_5 \leftrightarrow \tau_1$ and
$\tau_4 \leftrightarrow \tau_2$  lead to solutions of \rf{hirospr}. It turns
out that such procedure leads to all soliton solutions of $Sp(3)$ (in fact
this generalizes to any $SU(2N)$ and $Sp(N)$ \ct{ACFGZ}).

For instance, the solution \rf{suntaugalois} possesses such symmetry for $y_1 =
y_2 \equiv y/2$. Therefore the corresponding solution for $Sp(3)$ is given by
\br
\t_j = 1 + y \cos\({\pi j \o 3}\) e^\G + \( {3 \o 16}  y^2 + (-1)^j z \)
e^{2\G}
 + {z\o 3} y \cos\({2 \pi j\o 3}\)  e^{3 \G}
+(-1)^j {z y^2\o 48} e^{4 \G}
\lab{sprgalois}
\er
for $j=0,1,2,3$. The masses are given by
\br
M_1^{\l =1} &=&{48 \o{\psi^2}}m  \, \, \, \, \,
\mbox{\rm for $z=0$ and $y \neq 0$
or $z \neq 0$ and $y=0$}\\
M_2^{\l =1} &=&{96 \o{\psi^2}}m \, \, \, \, \,
\mbox{\rm for $z, y \neq 0$}
\er

\end{document}